\documentstyle[preprint,aps]{revtex}

\newcommand{\be}{\begin{equation}}
\newcommand{\ee}{\end{equation}}

\begin{document}



\title{ Motion of Vacancies in a Pinned Vortex Lattice: Origin of the
      Hall Anomaly }
\author{ P. Ao }
\address{Department of Theoretical  Physics, Ume\aa{\ }University \\                
         901 87, Ume\aa, Sweden    }

\date{\today}

\maketitle

\begin{abstract}
Physical arguments are presented to show
 that the Hall anomaly is an effect of 
the vortex many-body correlation rather than
that of an individual vortex. Quantitatively,
the characteristic energy scale in the problem, 
the vortex vacancy formation energy, is obtained for thin films. 
At low temperatures a scaling relation between the Hall 
and longitudinal resistivities is found, with the power depending 
on sample details. Near the superconducting transition temperature and 
for small magnetic fields the Hall conductivity is found to be 
proportional to the inverse of the
magnetic field and to the quadratic of the difference between the measured and 
the transition temperatures.

\noindent
PACS${\#}$s: 74.60.Ge
\end{abstract}



The ubiquitous occurrence of the Hall anomaly in the mixed state of 
both conventional and oxide superconductors, the sign change of the 
Hall resistivity below the superconducting transition temperature
and the smallness of the Hall angle,
has defied a consistent explanation so far\cite{exp}.
A straightforward application of the Magnus force 
cannot explain this phenomenon. 
This failure leads to a frustration of questioning the Magnus force
as the only transverse force in  the vortex dynamic equation.
The transverse force has been subsequently modified 
into various forms\cite{theory}.
Following the general properties of a superconductor, on the other hand,
recent theoretical work on vortex dynamics have shown that 
there is no other transverse force than the Magnus force,
a result of the topological constraint and the momentum
conservation.\cite{ao1,thouless}
One can further show that the Magnus force 
is equivalent to the spectral flow process.\cite{ao3}
An apparent conflict between the theoretical reasonings and 
the experimental measurements naturally arises.

In the present letter we attempt to solve this puzzle by showing 
that the Hall anomaly can be understood based on the vortex
vacancy motion in a pinned vortex lattice, and 
emphasize that the anomaly is a property of the vortex 
many-body correlation rather than that of an individual vortex.
We will demonstrate that vacancies can have the lowest energy scale, 
and they dominate the thermal activation contributions 
to the vortex motion at low temperatures.
The present vacancy model for the Hall anomaly 
is also consistent with other measurement such as the Nernst effect.
In the following we present our arguments leading to the model, 
and discuss its predictions and validity conditions.
For simplicity, we will consider an isotropic s-pairing superconductor with
one type of charge carriers in two dimension.
In this situation vortices(or straight vortex lines)
can be viewed as point particles.\cite{ao2}

The vortex dynamic equation for a $j$-th vortex of unit length in the sample
takes the form of the Langevin equation identical to that of a
charged particle in the presence of a magnetic field: 
\be
   m \ddot{\bf r}_j = q \frac{\rho_s}{2} h \; 
     ( {\bf v}_{s,t} - \dot{\bf r}_j )
    \times \hat{z} - \eta \dot{\bf r}_j + {\bf F}_{p} 
    + {\bf f} \; ,
\ee
with an effective unit length mass $m$, 
a pinning force ${\bf F}_{p}$, a vortex viscosity $\eta$, and a
fluctuating force ${\bf f}$. 
In Eq.(1) $q = \pm 1$ is the vorticity, $h$ the Planck
constant, $\rho_s$  the superfluid electron number density at temperature $T$,
and $\hat{z}$ the unit
vector in $z$-direction.
If there is  a temperature gradient, the thermal force
${\bf F}_T = - s_\phi \nabla T$ should be added in 
at the right hand of Eq.(1),
with $s_\phi$ the entropy carried by a vortex. 
The term associated with the total superconducting electron 
velocity 
${\bf v}_{s,t} = {\bf v}_s + {\bf v}_{s,in}$ 
and the vortex velocity $\dot{\bf r}$ at the right side of Eq.(1)
is the Magnus force. 
Although ${\bf v}_{s,t}$ is due to all other vortices, here 
we split it into two parts, with ${\bf v}_s$ corresponding to 
the rearrangement of vortices due to the externally applied supercurrent 
and ${\bf v}_{s,in}$ accounting for the rest contribution describing 
the vortex interaction without external current. 
In the following we will assume ${\bf v}_s$ is small such that 
this splitting is valid.

It is evident that in the mixed state of any real superconductor
the many-body correlation between vortices and the pinning effect
cannot be ignored. The competition between them is the source of the rich 
static and dynamical properties of flux phases\cite{blatter,bhattacharya}. 
We will take the Abrikosov lattice as the known manifestation of the
many-body correlation for the starting point to advance our arguments.
The vortex pinning is also important in our reasoning, though
several quantitative results obtained below do not explicitly
depend on it. 
If there were no pinning for vortices,
the whole vortex lattice would move together under the influence of 
an externally applied current in the same manner as that of 
independent vortices. Hence one would get the same sign of the Hall 
resistivity in both superconducting and normal states.
In the presence of pinning centers in the sample as well as the edge
pinning,
the vortex lattice will be pinned down.
In such a situation the motion of the vortex lattice is made possible 
by various kinds of defect motions due to thermal fluctuations. 
We will argue below that at low temperatures
the dominant contribution to the motion is due to vortex vacancies, 
and the Hall anomaly occurs.

For two vortices separated by a distance $r$,
which is less than the effective magnetic 
screening length $\lambda_{\perp}=\lambda_L^2/d$($d<\lambda_{L}$,
$\lambda_{\perp}=\lambda_L$ if $d>\lambda_L$) but greater than 
$\xi_0$, the interaction potential is
$
   V_0(r) = 2 d ( {\Phi_0}/{4\pi \lambda_L })^2 \; 
         \ln({r}/{\xi_0}) $.\cite{blatter}
Here $\lambda_L^2 = m^{\ast} c^2/8 \pi \rho_s e^2  $ is 
the London penetration depth,
$m^{\ast}$ the effective mass of a Cooper pair,
$\xi_0$ the coherence length of the superconductor, 
and $d$ the thickness of the superconductor film.
The energy scale 
$\epsilon_0 \equiv d ( {\Phi_0}/{4\pi \lambda_L } )^2$ 
sets both the scale for the strength of vortex
interaction and the scale for the strength of a strong pinning center. 
The energy for a dislocation pair separated by a distance 
larger than the vortex lattice constant is given by  
$  V_d(r) = ({\epsilon_0}/{ 2\sqrt{3}\pi})\ln({r}/{a_0})$,\cite{brandt,blatter}
with $a_0$ an order of the the vortex lattice constant.
The energy scale $\epsilon_0 / 2\sqrt{3} \pi$ 
for the dislocation pair here is about 10 times smaller than 
$\epsilon_0$ for the vortex interaction and pinning centers.
It is energetically favorable to have dislocation pairs
in the lattice. Hence for temperature $T << \epsilon_0$ we can ignore the 
contribution from the vortices hopping out of pinning centers
and the 
creation of vortex-antivortex  pairs.
The vortex lattice is then effectively pinned down.
Because vacancies and interstitials can be viewed as the smallest dislocation
pairs\cite{friedel}, 
we immediately have the estimated energy scale for vacancy formation energy 
$\epsilon_{v}$ as, by putting $r \sim 2 a_0$ in $V_d(r)$,
\be
    \epsilon_v \sim 
           \frac{1}{ 2\sqrt{3} \pi } 
            \left(\frac{ \Phi_0 }{4\pi \lambda_L} \right)^2 d \; .  
\ee
This result is valid for an intermediate magnetic field $B$: 
$ H_{c1} < B < H_{c2}/2$.  
For thicker films the thickness in Eq.(2) will be replaced
by a crossover thickness $d_c$ due to the $z$-direction correlation, 
whose precise value is 
a complex and unknown function of
various parameters such as the magnetic field, the pinning, 
the temperature, and anisotropy.
In the case $d_c$ is finite, 
its estimation in the high magnetic field limit is as follows.
Ring type vacancy excitations are possible in thicker films.
Its energy scale is determined by the smallest ring, which should be
the size of the vortex lattice constant. In this case $d_c$ 
is an order of the vortex lattice constant, $d_c \sim a_0$.
However, because of the large anisotropy in the HTcS materials,
$d_c$ can be the order of the CuO layer spacing close to the superconducting
transition temperature $T_{c0}$.

It is clear from the above analysis
that vacancies and interstitials have the lowest excitation energy scales.
We note that the value in Eq.(2) 
is consistent with the variational and numerical
calculations\cite{frey}, with about a factor 2 smaller, and 
also with the estimation from the dislocation core energy\cite{chui}.
Using the shear modulus results for $B \leq H_{c1}$\cite{blatter} and 
$H_{c2}/2 < B \leq H_{c2}$\cite{brandt,blatter} we have obtained
the corresponding vacancy 
formation energies as $ (2/9\pi)^{1/2} (b_0/\lambda_L)^{3/2} 
     e^{-b_0/\lambda_L} \; \epsilon_0 $ 
with $b_0^2 = \Phi_0/B$, and $0.7/\sqrt{3}\pi (1-B/H_{c2})^2 \; \epsilon_0$, 
respectively.

Now we argue that the vacancy formation energy 
is even lower than that of an interstitial.
The experimental observations at low magnetic fields have shown
the abundance of vacancies comparing with interstitials.\cite{energy}
The natural explanation is that the
the vacancy formation energy is lower than that of interstitials,
therefore by thermal fluctuations vacancies have a higher density.
The theoretical calculations have also confirmed
the lower vacancy formation energy\cite{hill}.
This phenomenon of the vacancy formation energy is lower than that 
of interstitials has also been observed in other crystalline 
structures\cite{friedel}. 
We conclude that vacancies will dominate thermal fluctuation contributions 
to resistivities at low enough temperatures for 
low magnetic fields.\cite{comment}

We show next that in the pinned 
vortex lattice a vacancy  behaves as a vortex with a vorticity $ - q$ 
and an interstitial as a vortex with  $ + q$, respectively.
Let ${\bf u}$ be the displacement vector at position ${\bf r}$, 
with a point defect, vacancy or interstitial, 
at ${\bf r}_0$. According to Eq.(1)
the transverse force acting on the defect is, 
measured from the pinned perfect vortex lattice, 
\be
   {\bf F}_M^d = q \frac{\rho_s}{2} h  \; 
   \int d^2{\bf r} \; \delta\rho({\bf r}) \; 
     ( {\bf v}_s - \dot{\bf u} ) \times \hat{z}  \; .
\ee
Here the vortex density $\delta\rho$ deviated from 
a perfect lattice 
is determined by the dilatation $\nabla\cdot{\bf u}$: 
$\delta\rho = \nabla\cdot{\bf u}/ {S_0}$,
with $S_0$ the area of a unit cell in the vortex lattice.
By definition, 
\be
   \nabla\cdot{\bf u} = \mp \; S_0 \; \delta^2({\bf r} - {\bf r}_0)\; ,
\ee
with `$-$' for a missing vortex, a vacancy,  and `$+$' for an extra vortex, 
an interstitial. 
Using Eqs.(3,4), we have the desired transverse force 
on the defect as
\be
   {\bf F}_M^d = \mp q \frac{\rho_s}{2} h  \; 
     ({\bf v}_s - \dot{\bf r}_0 ) \times \hat{z} \; .
\ee
This is identical to the dynamics of a hole or a particle
in a semiconductor in the presence of a magnetic field,  with 
a pinned perfect vortex lattice as a filled valence band and a vacancy in real
space as a hole in the energy space. 
Eq.(5) shows that both a vacancy and an interstitial will
move along the direction of the applied supercurrent ${\bf v}_s$.
This implies that vortices defining vacancies move against the direction of 
${\bf v}_s$, a result of the many-body correlation and pinning.
This leads us to our main conclusion that at low enough 
temperatures the sign of the Hall resistivity is different from its sign in 
the normal state because of the dominance of vacancies.
Quantitatively, vacancies and interstitials 
may be considered as independent particles moving in the
periodic potential formed by the vortex lattice and a random potential
due to the residue effect of pinnings.
The potential height of the periodic potential as well as that of 
the random potential is an order
of $\epsilon_v$. 
Assuming the vacancy (interstitial) density $n_{v}(n_{i})$ 
in a steady state, the longitudinal resistivity is
\be
    \rho_{xx} = \frac{h}{2 e^2} \sum_{l=v,i}
             \frac{\eta_{l} \; {\rho_s  h }/{2} }
                      { \eta_{l}^2 + ({\rho_s  h }/{2})^2 } \;
                \frac{ n_{l} }{\rho_s }  \; ,
\ee
and the Hall resistivity 
\be
   \rho_{yx} =  \frac{ h}{2 e^2} 
            \sum_{l=v,i} q_l \frac{ ({\rho_{s} h}/{2})^2 }
                   { \eta_{l}^2 + ({\rho_s  h}/{2})^2 } \;
                \frac{ n_{l} }{\rho_s}   \; , 
\ee
with $q_v = - q$ and $ q_i = q$.
Here $\eta_{v,i}$ are the effective vacancy and interstitial viscosities,
related to their diffusion constants in the
periodic potential due to the vortex lattice by the Einstein relation between
the diffusion constant and the mobility.
It should be pointed out that  contributions of other
vortex motions to resistivities such as vortex-antivortex pairs, 
which are omitted here for their 
smaller activation probabilities, 
are additive to those of vacancies, 
and that the including of the normal
fluid (quasiparticle) contributions is straightforward\cite{ao4}.

Under the driving of a temperature gradient,
the effective thermal force felt by a vacancy
is opposite in sign to the force felt by an interstitial 
or a vortex in direction 
but equal in magnitude, ${\bf F}_T^v = + s_\phi \nabla T$.
This can be seen by repeating the demonstration from Eq.(3) to (5).
Then the Nernst effect due to vacancies has the same sign as that of 
vortices or interstitials. Therefore our model gives that
in the Hall anomaly regime there is no sign change for 
the Nernst effect, and furthermore, 
the Nernst effect is more pronounced because of the
additive contributions due to both vacancies and interstitials.
This is in agreement with the experimental 
observations\cite{ao4}.

Before exploring of consequences of Eqs.(6,7) 
we discuss the qualitative implications of the present model.
In the above picture,  to obtain a maximum contribution
of vacancies, we need  the vortex lattice to define vacancies and
a sufficiently strong 
pinnings to prevent the sliding of vortex lattice. 
The existence of a whole lattice structure is nevertheless unnecessary.
Sufficiently large local crystalline structures, 
like lattice domains, will be enough to 
define vacancies. Therefore vacancy-like excitations in
a vortex liquid state can exist, 
because of the presence of large local orderings. 
Whether or not this is also true for a vortex glass state 
depending on details. 
For example, a further lowering of temperature may quench a 
vortex system into a glass state with no local crystalline 
structure. 
Then vacancies will disappear and the sign 
of the Hall resistivity will change again. 
On the other hand, for a fixed temperature 
if the pinning is too strong, for example, the (random) pinning center density
is much larger than the vortex density, vortices will
be individually pinned down and the local 
lattice structure required 
for the formations of  vacancies and interstitials will be lost. 
This suggests that the Hall anomaly only exists
in a suitable range of pinnings and magnetic fields, that is,
for $B_l < |B| < B_u$ with the lower and upper critical fields 
determined by pinning as well as by temperature. 

Now we study the limiting cases of Eqs.(6,7).
At low temperatures the 
motions of vacancies and interstitials in the vortex lattice 
are thermal hopping: 
 $\eta_v = \eta_0 \; e^{a_v \; \epsilon_v /K_B T } $ and  
 $\eta_i = \eta_0 \; e^{a_i \; \epsilon_v /K_B T } $,
with $a_v, a_i$(presumably $a_v < a_i$)  
numerical factors of order unity and $\eta_0$ insensitive to temperature.
In this limit, the vacancy (interstitial)
 density $n_v= n_0 \; e^{ - b_v \; \epsilon_v / k_B T}$
 ($n_i= n_0 \; e^{ - b_i \; \epsilon_v / k_B T }$), with 
 $b_v= 1 (b_i > 1) $ 
for the thermally activated vacancies(interstitials) and
$b_v(b_i) =0$ for the pinning center induced vacancies(interstitials). 
In the following 
we further assume that $\eta_v, \eta_i >> \rho_s  h /2$, corresponding to
the Hall angle $|\tan\theta| = |\rho_{yx}/\rho_{xx} | << 1$ 
common in experiments. 
Under this assumption,
we obtain the Hall angle as
\[
   \tan\theta = - q \frac{\rho_s h }{ 2\eta_0}
      \frac{ e^{ -(2 a_v + b_v) {\epsilon_v}/{k_B T} }
           - e^{ -(2 a_i + b_i) {\epsilon_v}/{k_B T} } }
           { e^{-(a_v + b_v) {\epsilon_v}/{k_B T} }
           + e^{-(a_i + b_i) {\epsilon_v}/{k_B T} }    }
\]

\be
   = \left\{ \begin{array}{ll}
      - q \frac{\rho_s h}{2\eta_0}
    \frac{\gamma}{2} \frac{\epsilon_v}{k_B T} \; , & 
     k_B T \geq \epsilon_v \; . \\  
      - q \frac{\rho_s h}{2\eta_0}
      e^{- a_v {\epsilon_v}/{k_B T} } \; ,
      & k_B T < min\{1,\gamma\}\epsilon_v \; . \end{array} \right.
\ee  

Here $\gamma = 2 a_i + b_i - 2 a_v - b_v$.  
The high temperature limit $k_B T \geq \epsilon_v$ is achieved near 
superconducting transition temperature $T_{c0}$, but the thermal creation 
of a vortex-antivortex pair is still improbable, because the relevant energy
scale $\epsilon_0$ is about 10 times bigger than $\epsilon_v$.
In the low temperature limit 
both longitudinal and Hall resistivities vanish exponentially. 
We obtain a scaling relation between them as
\be
   \rho_{yx} = A \; \rho_{xx}^{\nu } \; ,
\ee
with $A = - q (\rho_s h/2\eta_0)^{b_v/(a_v + b_v)} 
          (2e^2 \rho_s/h n_0 )^{a_v/(a_v + b_v)}  $,  
and the power 
\be
   \nu = \frac{2 a_v + b_v}{a_v + b_v} \; ,
\ee
varying between 1 and 2, depending on the detail of a sample which determines 
the numerical factors $a_v$ and $b_v$. 
If all vacancies are produced by pinnings, we have $b_v = 0$ and $\nu = 2$.
In this case $A$ is independent of $B$ because $n_0$ is.
In the other limit, if all vacancies are produced by thermal activations, 
and if $ a_v << 1$, we have $b_v = 1$ and $\nu \simeq 1$. 
In this case $A$ will be independent of $B$ if if $\eta_0$ is.

Another useful quantity is the Hall conductivity 
$\sigma_{xy} = \rho_{yx} /( \rho_{xx}^2 + \rho_{yx}^2 )$.
Under the same assumption of $\eta_v,\eta_i >> \rho_s  h /2$
we obtain the Hall conductivity due to vacancies and interstitials, 
from Eqs.(6,7), as
\[
  \sigma_{xy} =  - q \frac{ 2e^2 }{h }\frac{ \rho_s }{n_0 }
                \frac{ e^{ -(2 a_v + b_v) {\epsilon_v}/{k_B T} }
                     - e^{ -(2 a_i + b_i) {\epsilon_v}/{k_B T} } }
                     { \left[ e^{-(a_v + b_v) {\epsilon_v}/{k_B T} }
                + e^{-(a_i + b_i) {\epsilon_v}/{k_B T} } \right]^2 }
\]
\be
             = \left\{ \begin{array}{ll}
      - q \frac{ 2e^2 }{h }\frac{ \rho_s }{n_0 }
    \frac{ \gamma }{4} \frac{\epsilon_v}{k_B T} \; , & 
     k_B T \geq \epsilon_v \; . \\  
      - q \frac{ 2e^2 }{h }\frac{ \rho_s }{n_0 }
                e^{ + b_v {\epsilon_v}/{k_B T} }\; ,
      & k_B T < min\{1,\gamma\}\epsilon_v \; . \end{array}\right. 
\ee    
As discussed above, here $0 \leq b_v \leq 1$ and 
$\gamma \sim O(1)$.
Near the superconducting transition temperature $T_{c0}$, $\rho_s 
= \rho_{s0} ( 1 - T/T_{c0})$ and $\epsilon_v = \epsilon_{v0}( 1 - T/T_{c0})$
because of the London penetration depth in Eq.(2).
We may further assume $n_0 = B/\Phi_0$, with $\Phi_0$ the flux quantum.
From Eq.(11) we obtain  
\be  
   \sigma_{xy} = \alpha_1 \frac{ ( 1 - {T}/{T_{c0} })^2 }{B} \; ,   
\ee
with 
$ \alpha_1 = - q ({ 2e^2 }/{h }) \rho_{s0} \Phi_0 
    \gamma {\epsilon_{v0} }/{4 k_B T_{c0}} $.
Taking $\rho_{s0} = 10^{21} /cm^3$,
$\gamma = 1$, and ${\epsilon_{v0} }/{k_B T_{c0} } = 50$,
we find $|\alpha_1 |\sim 20 \, T\mu\Omega^{-1}cm^{-1}$.

Two comments are in order.
1. A naive accounting of 
the many-body correlation and pinning may not lead to the sign change in the
Hall resistivity:
Since vortex interaction terms cancel each other when summing 
over all vortices, one would like to conclude that there is no
many-body correlation effect on the sign of the Hall resistivity.
If this claim were correct, the same argument  
would lead to no sign change
for the Hall resistivity in a hole semiconductor, and no such phenomena as
the quantum Hall effect.
This absence of Hall anomaly is the result of  the
underestimation of the many-body correlation.
2. There might be a tendency to mix up antivortices and vacancies.
As discussed above,
the creation energy of an antivortex is about 10 times larger 
than that of a vacancy, 
which makes it energetically unfavorable. Furthermore, since
an antivortex feels the same thermal force as a vortex,
${\bf F}_T = - s_\phi \nabla T$, it has an opposite sign  contribution to 
that of a vortex for the Nernst effect, in conflicting with 
experiments.\cite{exp}

In conclusion, we have demonstrated 
that within the vortex dynamics equation the Hall anomaly 
can be explained.
What has been missed in previous models is a proper consideration of the
competition between the many-body correlation and pinning.
We have proposed the model of vacancy  motion in a pinned lattice
as a concrete realization:
the characteristic energy in the model, the vacancy formation energy, 
is obtained; and 
vacancies move along an applied supercurrent as the origin for the Hall 
anomaly.
The model leads to an exponential tail and the scaling relation
at low temperatures, and no sign change for the Nernst effect.
Near the superconducting transition temperature and for small magnetic fields
the Hall conductivity is found
to be proportional to the inverse of the magnetic field and is quadratic 
in the temperature different from the transition.
For thin enough films the activation energy in the low temperature limit
has a linear film thickness dependence. 

\noindent
{\bf Acknowledgments} 
 Helpful conversations with L. Gor'kov, P. Minnhagen, and especially with 
 D. Thouless are gratefully acknowledged.
 The hospitality of Physics Department at University of Washington,
  where part of the work was done, is also appreciated.
 This work was supported in part 
 by Swedish Natural Science Research Council and by US NSF grant no. 
 DMR-9220733.



\begin{thebibliography}{99}
\bibitem{comment}
 In the high magnetic field limit (logarithmic interaction) 
 the numerical calculation  in Ref.[12] has shown that 
 interstitials have lower formation energy instead, which may explain the 
 disapparency of the Hall anomaly in that limit.
\end{thebibliography}
\end{document}